\documentclass[letterpaper]{jpconf}
\usepackage{citesort}
\usepackage{graphicx}
\begin{document}
\title{The high-field Fermi surface of YbRh$_2$Si$_2$}

\author{P~M~C Rourke$^{1}$, A McCollam$^{1}$, G Lapertot$^{2}$, G Knebel$^{2}$, J Flouquet$^{2}$ and S~R Julian$^{1}$}
\address{$^{1}$Department of Physics, University of Toronto, Toronto, ON M5S 1A7, Canada}
\address{$^{2}$DRFMC, SPSMS, CEA Grenoble, 17 rue des Martyrs, 3805-4 Grenoble cedex 9, France}

\begin{abstract}

We present quantum oscillation measurements of YbRh$_2$Si$_2$ at magnetic fields above the Kondo-suppression scale $H_{0} \approx 10$~T. Comparison with electronic structure calculations is complicated because the ``small'' Fermi surface, where the Yb 4$f$-quasi-hole is not contributing to the Fermi volume, and ``large'' Fermi surface, where the Yb 4$f$-quasi-hole is contributing to the Fermi volume, are related by a rigid Fermi energy shift. This means that spin-split branches of the large Fermi surface can look like unsplit branches of the small surface, and vice-versa. Thus, although the high-field angle dependence of the experimentally-measured oscillation frequencies most resembles the electronic structure prediction for the small Fermi surface, this may instead be a branch of the spin-split large Fermi surface.

\end{abstract}

The heavy fermion compound YbRh$_2$Si$_2$ has been the focus of much recent interest because it can be tuned through a field-induced quantum critical point (QCP) that is both easily accessible to experiment and appears to exhibit a new class of ``local'' quantum criticality~\cite{gegnaturephysreview}. Such a scenario implies different Fermi surface (FS) topologies in various regions of the YbRh$_2$Si$_2$ phase diagram~\cite{gegnaturephysreview}, some of which can be probed directly via quantum oscillation measurements.

\begin{figure}
\begin{center}
 \includegraphics[width=0.95\columnwidth]{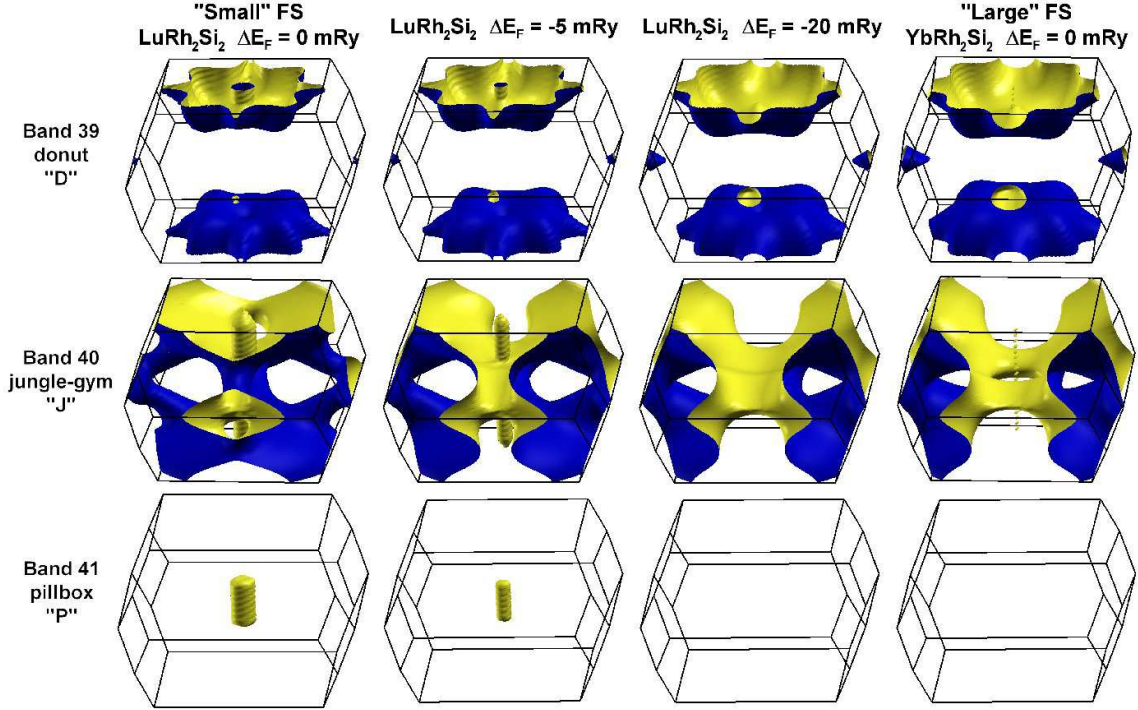}
 \caption{\label{fig:fssheets} ``Small'', ``large'', and two intermediate Fermi surfaces calculated with the LDA + spin-orbit coupling method. Following convention~\cite{wigger}, the sheets are labelled donut ``D'', jungle-gym ``J'', and pillbox ``P''. From an electron point of view, the dark blue side of each sheet is the occupied side and the light yellow side is the unoccupied side, such that D is a large hole surface and P is a small electron surface.}
 \end{center}
\end{figure}

The two proposed Fermi surface topologies are denoted ``small'' and ``large'': In the small FS case (also known as ``4$f$-localized'' or ``Yb$^{3+}$''), the Yb 4$f$ quasi-hole does not contribute to the Fermi volume; whereas in the large FS case (also known as ``4$f$-itinerant'' or ``Yb$^{2+}$''), the Yb 4$f$ quasi-hole does contribute to the Fermi volume. The Fermi surfaces corresponding to these two configurations are shown in Fig.~\ref{fig:fssheets}. At $T = 14$~K and zero magnetic field, below the Kondo scale $T_{K} \approx 25$~K, the band structure corresponding to the small FS case is observed by ARPES measurements~\cite{wigger}. As the temperature is decreased, non-Fermi-liquid behaviour is observed down to the phase transition into a weak antiferromagnetic (AF) state at $T_{N} \approx 70$~mK~\cite{trovarelli}. As the magnetic field is increased at low temperature, the weak magnetism is suppressed, leading to a QCP at $H_{c} \approx $ 0.06~T for $H \bot c$~\cite{gegearlyprl}. For fields above the QCP, a region of Landau Fermi liquid behaviour is recovered~\cite{gegnjp}.

A line of Hall coefficient crossovers has been discovered in the $H$--$T$ phase diagram, extrapolating to a discontinuous jump at the QCP. This has been interpreted as signalling a sudden Fermi surface reconstruction, suggested to be from the small FS (for $H < H_c$) to the large FS (for $H > H_c$)~\cite{paschennature}. The additional association of crossovers in numerous other physical quantities with the Hall coefficient crossover line, implying a second energy scale at the QCP~\cite{gegscience}, has led to the proposal of a new type of ``local'' quantum criticality involving the breakdown of Kondo screening at the QCP~\cite{gegnaturephysreview}. As the field is further increased at low temperature, the Kondo physics is suppressed by field at $H_{0} \approx 10$~T. The Yb 4$f$ state has been thought to behave as a local moment for $H > H_{0}$, and indeed measurements of dc-magnetization, $T^{2}$ coefficient of resistivity, specific heat, and the linear magnetostriction coefficient have been interpreted as evidence for a continuous crossover from a large FS to a small FS across $H_{0}$~\cite{gegnjp}.

Until recently, however, there was no direct experimental evidence for the size or shape of the Fermi surface at 0~K in any of these regimes.  This situation improved when some of us~\cite{knebel} observed quantum oscillations well above 10~T for magnetic fields applied in the basal plane. In principle, such studies can distinguish between the large and small Fermi surface scenarios by comparing the measurements with electronic structure calculations corresponding to both cases. Since Lu and Yb differ by exactly one 4$f$ hole, an all-itinerant calculation of isostructural LuRh$_2$Si$_2$ yields a Fermi surface nearly identical to the YbRh$_2$Si$_2$ small FS (that is, without the 4$f$ quasi-hole contributing to the Fermi volume)~\cite{wigger}, whereas an all-itinerant calculation of YbRh$_2$Si$_2$ results in the large FS. We have performed these calculations using the WIEN2k density functional theory code, which is based on the all-electron full-potential augmented plane-wave + local orbitals method, and includes relativistic effects and spin-orbit coupling (SOC) at the one-electron level~\cite{wien2k}. A tetragonal unit cell with space group $I4/mmm$ (\#139), lattice constants $a = 4.010$~\AA~and $c = 9.841$~\AA, and relative atomic positions Yb/Lu (0,0,0), Rh (0,0.5,0.25), and Si (0,0,0.375) was employed. Band energies were obtained on a 20~000 k-point mesh, using the Perdew-Burke-Ernzerhof generalized gradient approximation to the exchange-correlation potential~\cite{pbe96}, $R~K_{max} = 8$, and an energy range of $-7.1$ to 5~Ry. Our final calculations did not include an intra-atomic repulsion $U$, since this had a negligible effect on the Fermi surfaces. The resulting small Fermi surface, visualized in Fig.~\ref{fig:fssheets} via the XCrysDen program~\cite{xcrysden}, is indistinguishable from that published previously~\cite{wigger}; the band structure and density of states vs.\ energy giving rise to the large Fermi surface (also shown in Fig.~\ref{fig:fssheets}) also agree in detail with previously-published all-itinerant LDA+SOC calculations of YbRh$_2$Si$_2$~\cite{jeong}. Predicted dHvA frequencies were extracted from the calculated Fermi surface sheets of Fig.~\ref{fig:fssheets} using an algorithm described elsewhere~\cite{skeaf}.

Within these calculations, under a rigid Fermi energy shift of approximately $-30$~mRy, the small FS looks very similar to the large FS. Fermi surfaces for two intermediate cases (the small FS with $E_F$ shifts of $-5$~mRy and $-20$~mRy) illustrating this transformation are shown in Fig.~\ref{fig:fssheets}. Similarly, under a shift of approximately $+20$~mRy (not shown), the calculated large FS looks like the small FS, raising the possibility that under sufficiently-strong spin-splitting of the bands at high magnetic fields one of the spin-branches of the large FS could look like the small FS, without requiring a localization transition. Furthermore, a Fermi energy shift of approximately $-50$~mRy (not shown) transforms the donut ``D'' sheet of the large FS case into a sheet topologically similar to the large FS jungle-gym ``J'' sheet, revealing a band shape resemblance between the ``D'' and ``J'' bands. Note that the fact that the energy of these shifts is very large is not a concern: these are the shifts given by LDA, which greatly overestimates the dispersion of the bands, typically by factors of several hundred in heavy fermion materials.

The previous study, using a torque method, although confined to the basal plane, suggested that the small Fermi surface scenario gives better agreement with the measured quantum oscillation frequencies~\cite{knebel}.  Here we extend this work by using the standard field modulation technique.  This has allowed us to follow the signals toward the c-axis in the (110)--(001) plane. The high-quality single-crystal samples studied were $\sim 0.2 \times 1 \times 2$~mm platelets grown from In-flux~\cite{knebel}, with a residual resistivity ratio of about 100. Measurements between 14--16~T, below 50~mK, with magnetic field rotated in-situ across a range 45 degrees from $a$ = (100) to (110) and 60 degrees from (110) toward $c$ = (001), were performed in a specially-designed cryomagnetic facility composed of a $^3$He/$^4$He dilution refrigerator, superconducting magnet and dedicated modulation coils.

\begin{figure}
\begin{center}
 \includegraphics[width=0.49\columnwidth]{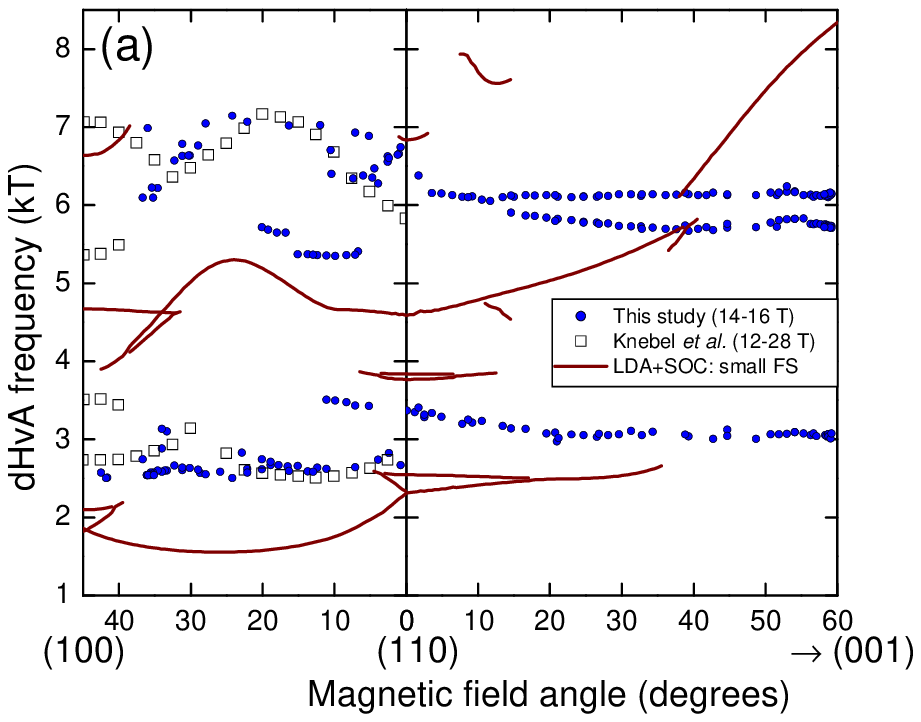}
 \includegraphics[width=0.49\columnwidth]{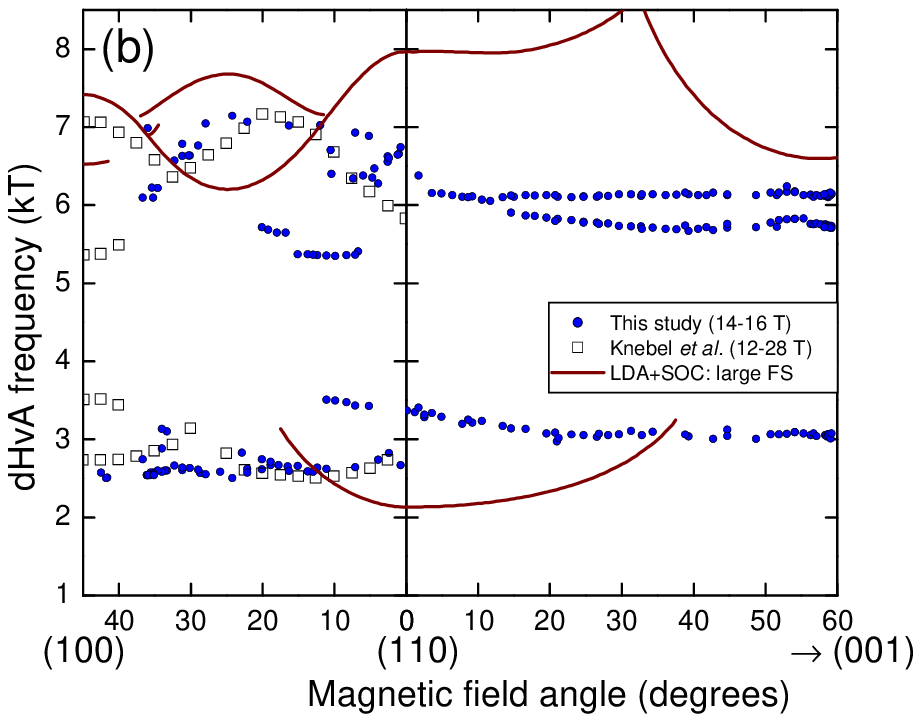}
 \caption{\label{fig:lufreqvsang} de Haas--van Alphen frequency vs.\ magnetic field angle. Closed blue circles depict the current experimental data (14--16~T field range), open squares show previously-published data (12--28~T field range)~\cite{knebel}, and red lines are from the calculated small (a) and large (b) Fermi surfaces shown in Fig.~\ref{fig:fssheets}.}
 \end{center}
\end{figure}

Fig.~\ref{fig:lufreqvsang} shows the field-angle dependence of our measured quantum oscillation frequencies. For comparison, previously-published data from torque measurements on samples grown in the same facility (in a 12--28~T field range)~\cite{knebel}, as well as predicted dHvA frequencies from the calculated small (Fig.~\ref{fig:lufreqvsang}a) and large (Fig.~\ref{fig:lufreqvsang}b) FS's of Fig.~\ref{fig:fssheets}, are included. With the exception of the extra frequency near 6~kT beyond $\sim 15^{\circ}$ in the (110)--(001) plane, and persistence of observed frequencies over a larger angular range in this plane, there is good qualitative agreement between the small FS calculation and experiment. Qualitative agreement between electronic structure calculations and quantum oscillation experiments on the FS topology (manifested by the frequency angle-dependence), but not quantitative agreement on the FS size (manifested by the frequency magnitudes), is usual for heavy fermion compounds, since, as seen in Fig.~\ref{fig:fssheets}, the flatness of the bands makes the calculated Fermi surface sizes depend sensitively on small shifts of the Fermi energy. The qualitative agreement between the non-spin-split large FS calculation and experiment is significantly poorer, since the calculation misses numerous branches in the (100)--(110) plane.

While the angle dependence of our measured de Haas--van Alphen frequencies appears to strengthen the belief~\cite{gegnjp} that for $H > H_0$ the small Fermi surface is realized, our investigation of the band structures shows that this could equally be a branch of the large Fermi surface that has undergone large spin-splitting. This finding points to the need for more realistic treatments of spin-splitting within electronic structure calculations.

\ack

The authors acknowledge useful discussions with S.~E. Sebastian. This work is supported by the Natural
Science and Engineering Research Council of Canada and the Canadian Institute for Advanced Research.

\section*{References}
\bibliography{LT25YbRh2Si2}

\end{document}